\documentstyle[12pt]{article}

\newcommand{\be}{\begin{equation}}
\newcommand{\ee}{\end{equation}}
\newcommand{\bea}{\begin{eqnarray}}
\newcommand{\eea}{\end{eqnarray}}

\newcommand{\g} {g}
\newcommand{\p} {\phi}
\newcommand{\V} {V}
\newcommand{\ggg} {{\overline g}}
\newcommand{\ppp} {{\overline \phi}}

\begin{document}

\begin{center}
\begin{large}
{\bf  Quantum-Corrected  Entropy \\}
{\bf  for \\}
{\bf  1+1-Dimensional Gravity  \\}
{\bf Revisited \\}
\end{large}  
\end{center}
\vspace*{0.50cm}
\begin{center}
{\sl by\\}
\vspace*{1.00cm}
{\bf A.J.M. Medved\\}
\vspace*{1.00cm}
{\sl
Department of Physics and Theoretical Physics Institute\\
University of Alberta\\
Edmonton, Canada T6G-2J1\\
{[e-mail: amedved@phys.ualberta.ca]}}\\
\end{center}
\bigskip\noindent
\begin{center}
\begin{large}
{\bf
ABSTRACT
}
\end{large}
\end{center}
\vspace*{0.50cm}
\par
\noindent

In this paper, we examine a generic theory of
1+1-dimensional gravity with coupling to a 
scalar  field. Special attention is paid to a class
of models that have  
 a power-law form of  dilaton potential and
can capably admit  black hole solutions.
The study  focuses on  the formulation of
a Lorentzian partition function.  We 
incorporate the principles of Hamiltonian thermodynamics,
as well as  black hole spectroscopy,
and find that the partition function can be expressed
in a well-defined,  calculable form. We then go on to
extract the black hole entropy, including the leading-order
quantum correction.  As anticipated,  this correction can
be expressed as the logarithm of the classical entropy.
Interestingly, the prefactor for this  logarithmic correction
disagrees, in both  magnitude and sign, with  the findings 
from a prior study (on the very same model).
We comment on this discrepancy and provide a possible
rationalization.

\newpage

\par

\section{Introduction}

There is a long and storied history to the branch of gravitational
physics known as black hole thermodynamics. It began in
the mid-seventies, when Bekenstein \cite{bek} and Hawking \cite{haw} 
formulated a convincing  analogy between black hole
mechanics and the laws that govern a thermodynamic system. 
Hawking dramatically punctuated this analogy with his
demonstration that black holes  emit thermal radiation
(via a quantum tunneling mechanism)  at a temperature
that complies perfectly with the  value predicted by
the  first law \cite{HAW2}. 
\par
There has since been an ``endless'' series of research  
papers that have examined
black hole thermodynamics from a myriad of different
angles \cite{WALDY}. 
The  motivation to study   black holes  has  stemmed
from several sources; for instance,
 their
likely existence  as astronomical objects, the role
of microscopic holes  in the  primordial universe, as well as 
simple aesthetic interest. However, the most compelling
motivating factor  might be  in the context of quantum gravity,
where black holes have served as a particularly 
useful testing ground  for theoretical frameworks and ideas.
In this regard, black hole entropy occupies a very  esteemed
position, as it is often argued that any viable theory
of quantum gravity must  naturally incorporate and ultimately
explain the Bekenstein-Hawking  area law \cite{bek,haw}.
\par
This important role of black hole entropy 
also  serves
to magnify the conceptual difficulties  
that it unfortunately  presents. Most significantly,
the information loss paradox \cite{PRES}, the  breakdown
of quantum field theory (or, equivalently, the holographic
storage of information  \cite{THO}), and the statistical
origin of the entropy in question \cite{carlip2}.  
That is not to say  there has been  a lack of progress in 
answering these questions; if anything, there has been
too much! For instance, various microstate-counting
frameworks  have successfully reproduced the
Bekenstein-Hawking entropy; including those with their 
origins in
 string theory \cite{str2}, quantum geometry \cite{ash},
Sakharov's induced gravity \cite{fro}, Chern-Simons theory \cite{carlip3},
asymptotic conformal symmetries \cite{str}, {\it etcetera}.
It is, however,  conspicuously unclear as to how these various techniques
are related and, moreover, as to what degrees of
freedom are truly being counted. Without some fundamental
theory that ties this altogether, the statistical origin
of the entropy remains as enigmatic as ever.
\par
It should be  evident  that we need to better our understanding of black hole
entropy;
however, 
it remains quite unclear as to how substantial progress can be made.
One possible direction, which has
enjoyed success in the past,   is  to investigate
theoretical models  of two-dimensional  gravity \cite{NEW1,KUM}.
Such  models  allow one to focus  directly on the 
essence of the physical issues, without being bogged
down by the calculational complexities inherent to
higher-dimensional theories. 
\par
Although the physical significance of  two-dimensional gravity
can, in general, be challenged, there are many
specific instances  where a given  model is  directly 
related to a physically relevant theory.
As an example, let us consider Jackiw-Teitelboim theory \cite{jt}, which
describes constant-curvature black hole solutions  in
 two-dimensional anti-de Sitter space. This model 
happens to have a   dual relationship  with
certain string-inspired black holes \cite{youm} and the
near-extremal sector of the Reissner-Nordstrom black hole \cite{ext}.
Moreover, the Jackiw-Teitelboim model can
also be viewed as a dimensionally reduced form of
the always topical BTZ black hole \cite{btz,ao}. 
\par 
With the above discussion in mind, the current paper
considers the black hole thermodynamics  of  a  general model
of 1+1-dimensional gravity  coupled to a dilatonic scalar field.
One of the principal achievements will be the  formulation of
a Lorentzian partition function, which  is inspired
by the  Hamiltonian thermodynamic analysis of Kuchar \cite{KUCHAR}
and others 
(for instance, \cite{LW,BPP,kun2}).  
To express the partition function in a  suitable form
for  explicit  analysis, we also call upon
some other intriguing  facets of black hole quantum theory; including
 spectroscopy \cite{BEKX},  complementarity
\cite{COMP} and  spacetime Euclideanization 
 \cite{HH}. In this sense, our formalism
provides a   simplified (lower-dimensional) framework
for illustrating some of the essentials of black hole
physics.
\par
Our  ultimate objective  will be to evaluate 
the black hole entropy, including the leading-order 
quantum correction to the classical thermodynamic value.
In this regard, it is interesting to take note
of a   related work by the current author \cite{MEME}. 
 This recent paper considered the quantum-corrected  entropy
 for precisely the same  two-dimensional
model. However, the  approach of  \cite{MEME}
was  very much different from that of the  present treatment.
In fact,  \cite{MEME} was based on  two distinct
analytic methods for  calculating the entropy; with
these being as follows:
\par
 Firstly, 
we  employed   a formula that was derived by Das {\it et al}
\cite{das} and follows from purely thermodynamic principles.
Secondly, we applied  Carlip's   quantum-corrected
version \cite{carlip} of the renowned Cardy  formula 
\cite{car}.\footnote{In using the  Cardy-Carlip formula, 
we first demonstrated
that the near-horizon geometry of the  generic
theory  can be effectively described by a conformal field
theory. Moreover, this conformal theory was shown to have an 
 identifiable Virasoro algebra.  This portion
of the analysis generalized a treatment by Solodukhin \cite{sol}.}  
\par
For both of these prior methods, the  leading-order 
 correction
was shown to be proportional to the logarithm 
of the classical entropy.  This  outcome is in agreement
with  various calculations
that are found  in the literature;  for instance,
in the quantum-geometry  analysis of Kaul and Majumdar \cite{kaul}.
(Also see, for example,  
\cite{dim}-\cite{med},\cite{carlip},\cite{blah0}-\cite{blah3},
\cite{das}, \cite{NEW3}, \cite{NEW4}.)
However, there was also a notable discrepancy between
the  two methods of \cite{MEME}: the statistical (i.e., Cardy)  
approach consistently yields a logarithmic prefactor of $-3/2$,
whereas the thermodynamic treatment
generates a model-dependent prefactor. Interestingly,
the two prefactors are only in agreement  for  the very special case
of Jackiw-Teitelboim theory \cite{jt}. 
\par
It should be noted
that a logarithmic prefactor of $-3/2$  
coincides precisely with  the value prescribed by quantum geometry \cite{kaul}.
This particular  value turns up elsewhere in the literature,
although not exclusively so. For instance, a four-dimensional study 
that is 
close in spirit to the current two-dimensional analysis  found a
logarithmic prefactor of $+1/2$ \cite{MR}.
One might  anticipate that the same outcome  will be found here;
but, as always in physics,  there  
can be  no substitute for an explicit check.
\par
The remainder of the paper is organized as follows.
In  Section 2, we introduce a  generic model of
  1+1-dimensional gravity with coupling to a scalar field. 
  We then go on to
 discuss the  black hole solutions  and associated thermodynamics
at the classical level. 
In Section 3, we consider a Lorentzian  partition function  for 
the 1+1-dimensional black hole of interest.
With the help of 
 Hamiltonian  thermodynamics and black hole spectroscopy,
we  are able to  express the partition function  
in a convenient form for explicit analysis. 
In Section 4, we  use the prior derivation  
as the basis for a calculation of the black hole entropy.
With the assumption of a semi-classical regime, we  are able to 
express the entropy  in terms of a well-defined
perturbative expansion.
Section 5 contains a summary and overview. In particular,
we directly compare the current results 
with the findings from  our prior study.

\section{Generic Dilaton Gravity}

Let us begin by considering a 1+1-dimensional theory of gravity
coupled to a dilaton (or auxiliary scalar) field.\footnote{The
dilaton is a necessary element, inasmuch as the Einstein
tensor identically vanishes for two-dimensional gravity.}
Assuming a diffeomorphism invariant action that contains
at most second derivatives of the fields, we have:
\be
I={1\over 2G}\int d^2x \sqrt{-\ggg}\left[D(\ppp)R(\ggg)+{1\over 2}
\ggg^{\mu\nu}\nabla_{\mu}\ppp\nabla_{\nu}\ppp +{1\over l^2} U(\ppp)\right].
\label{1}
\ee
Here, $G$ is a dimensionless measure of gravitational coupling
(i.e., the two-dimensional ``Newton constant''), $l$ is a 
fundamental constant of dimension length,
while $D(\ppp)$ and $U(\ppp)$ are well-behaved but otherwise
 arbitrary functions of the
dilaton field. 
It is a point of interest that the very same action can also describe   
 dilaton-gravity 
 coupled  to an Abelian gauge field.\footnote{Assuming 
 a gauge-invariant action, one finds that  the Abelian 
sector  can 
  be exactly solved in terms of  the dilaton and a conserved charge 
\cite{NEW2,abelian1}. Hence, the  
 total action can always be re-expressed in the form
of Eq.(\ref{1}).} 
\par
The viewpoint of this paper will be  that Eq.(\ref{1})
represents some sort of fundamental theory. Nevertheless,
 it is worth noting the above action
can often have  physical significance from the perspective
of  higher-dimensional
theories. For instance, one finds that $D=\ppp^2/4$ and $U=1$ 
after the spherical reduction of
   3+1-dimensional Einstein gravity  \cite{spher}.
Furthermore,  many  two-dimensional theories  
have  near-horizon dualities  with   
either  near-extremal black holes \cite{ext} or  string-theoretical models
\cite{youmy}.
\par
Exploiting conformal symmetry, one can
conveniently reformulate the  action 
so that
 the kinetic term is  eliminated. This task can, in fact,
be accomplished by way of the following field redefinitions
 \cite{kun}: 
\be
\p=D(\ppp),
\label{3}
\ee
\be
\g_{\mu\nu}=\Omega^2(\ppp)\ggg_{\mu\nu},
\label{2}
\ee
\be
\Omega^{2}(\ppp)=\exp\left[{1\over 2}\int^{\ppp}d\xi\left({dD\over d\xi}
\right)^{-1}\right],
\label{4}
\ee
\be
\V(\p)={U(\ppp)\over\Omega^{2}(\ppp)}.
\label{5}
\ee
In the above usage, it is implicit that
  $D(\ppp)$ and its derivative are  non-vanishing
throughout the relevant manifold. 
\par
In its reparametrized form, the action (\ref{1}) 
simplifies as follows:
\be
I={1\over 2G}\int d^2x \sqrt{-\g}\left[\p R(\g)+
{1\over l^2} \V(\p)\right].
\label{6}
\ee
\par
The field equations of this reparametrized action can 
readily be  solved. Moreover,
for the submanifold $x\geq0$,
 this
 solution can  be conveniently
expressed in a static, ``Schwarzschild-like'' gauge \cite{kuny,NEW2}:
\be
\p=\p(x)={x\over l} \quad \geq 0, 
\label{7}
\ee
\be
ds^2=-\left(J(\p)-2lGM\right)dt^2+ \left(J(\p)-2lGM\right)^{-1}dx^2,
\label{8}
\ee
\be
J(\p)=\int^{\p}d\xi\V(\xi),
\label{9}
\ee
where $M$  
is a constant of integration. We will regard $M$ as being
non-negative; in which case, it can  be identified
with the conserved mass of a black hole solution (if such a solution
 exists).
\par
In the analysis to follow, we will assume that the theory always admits
black hole solutions for which the outermost horizon, $\p_o=x_o/l$,
is   non-degenerate.\footnote{It is sufficient criteria for
a black hole solution if $J(\phi)-2lGM$ has a zero and ${dJ\over d\phi}>0$
 in the relevant manifold
($\phi\geq 0$). 
For a more rigorous discussion,  see \cite{CADY1}.}
In very general terms, a  black hole horizon
can be identified with 
a hypersurface of vanishing Killing vector \cite{wald}. For the
model under consideration,
 this identification  translates into the following relation \cite{mann,kuny}:
\be
J(\p_o)-2lGM=0.
\label{10}
\ee 
Alternatively, one can write:
\be
\p_o=J^{-1}(2lGM).
\label{100}
\ee
\par
Next, let us consider the tree-level black hole thermodynamics.
It is pertinent to this discussion that  all thermodynamic
properties of current interest are invariant under the
above reparametrization \cite{CADY2}.
Firstly, we can calculate the
Hawking temperature  ($T_o$)  via   the usual Euclidean
prescription \cite{HH}; that is, the inverse temperature
is identified with the periodicity of
imaginary  time.  One readily finds that:
\be
T_{o}={1\over 4\pi l}\left. dJ\over d\p\right|_{\p_o}.
\label{11}
\ee
\par
Given a two-dimensional spacetime, there is  no obvious way
of defining the area of a one-dimensional surface, such
as the horizon area  of a black hole.
For this reason,
the Bekenstein-Hawking area  law \cite{bek,haw} can not be exploited
in a straightforward manner.
Nonetheless, we can still
ascertain  the  thermodynamic entropy ($S_o$) by 
virtue of  the first law
of thermodynamics: $dM|_{horizon}=T_{o}dS_{o}$.  Integration of this
expression yields:
\be
S_{o}={2\pi\over G}\p_o,
\label{12}
\ee
where the arbitrary constant  has been set to zero
in accordance with the usual convention.
Working backwards from this result,
we can now define an effective horizon ``area''
as follows:
\be
A_o= 4G S_{o} =8\pi\p_o.
\label{area}
\ee
Here,  we have simply  applied the standard Bekenstein-Hawking  
definition.
\par
So far, we have treated the
 dilaton potential, $U(\ppp)$ or $\V(\p)$,  as   generically
as  possible. 
However, for illustrative
purposes, it will often prove to be instructive
and convenient
 if this potential is given an
explicit form.  Following our prior, related work \cite{MEME},
we thus consider a ``power-law''  potential:
\be
\V(\p)= \gamma\p^a ,
\label{15}
\ee
where $a$ and $\gamma$ are dimensionless, non-negative, model-dependent
 parameters.
Interestingly,
this form  of potential can   describe a 
Weyl-rescaled CGHS model (for $a=0$ and $\gamma=1$) \cite{cal, cad2}, or
a dimensionally reduced
 BTZ black hole (for $a=1$ and $\gamma=2$) \cite{btz,ao}.
On a more general note, such a potential can
capably  describe (after suitable rescalings)   
 the  near-horizon geometry of a single-charged
 dilatonic black hole, a multi-charged stringy black hole, or
a dilatonic $p$-brane \cite{youmy}.
Also, if  $a$ is allowed to be less than zero, than this potential
can  describe  the
spherical reduction of $d$-dimensional   Einstein  gravity
(for   $a=-1/(d-2)$ and $\gamma=(d-3)/(d-2)$
\cite{kun2}).  A negative value of $a$, however, somewhat complicates
the  later analysis.\footnote{This complication can be viewed
as a manifestation of such  theories admitting black holes with
a negative specific heat. The very same problem, of course,
 notoriously afflicts
the Schwarzschild black hole.}
Hence,  we will generally enforce $a\geq 0$ and
 comment further on this restriction at an appropriate  point. 
 \par
Given the power-law form of the potential (\ref{15}),
it is useful to note that (cf. Eqs.(\ref{9},\ref{100})):
\be
J(\p)={\gamma\over a+1} \p^{a+1},
\label{150}
\ee
\be
\p_o=\left[{a+1\over\gamma}2lGM\right]^{1\over a+1}.
\label{1500}
\ee
We can also   
re-express the black hole mass (\ref{10}) and temperature (\ref{11})
 as follows:
\be
M={1\over 2lG}{\gamma\over a+1} \p_o^{a+1},
\label{16}
\ee
\be
T_{o}={\gamma\over 4\pi l}\p_o^a,
\label{17}
\ee
whereas the entropy  only depends implicitly (through $\phi_o$)
on the specific choice of model.

\section{Hamiltonian Partition Function}

In this section, we will consider the  Lorentzian partition function
for a 1+1-dimensional (dilatonic) black hole. Formally speaking, 
this partition function can be defined as follows:
\be
{\cal Z}[\beta,\phi_{+}]=Tr\left[\exp(-\beta {\hat H})\right].
\label{301}
\ee
Here, $\phi_+$  is the fixed value of the dilaton at a
 timelike outer boundary of the system,
$\beta^{-1}$ is the  equilibrium temperature of the system
(as measured at the outer boundary), ${\hat H}$ is the Lorentzian 
Hamiltonian operator (discussed  below), and the trace
is over all physically relevant states.
\par
The premise of the above formulation is that a black hole can be 
viewed as thermodynamic object in a heat bath  of
fixed temperature $\beta^{-1}$ \cite{HH}.  
It is implied that 
the entire system has been  enclosed in a thermally reflective
``box''; thus  ensuring  that equilibrium is 
maintained.\footnote{Eventually,  we consider the asymptotic
limit of an infinitely large box or $\phi_+\rightarrow\infty$.}
The above partition function then appropriately describes 
the thermodynamics of a
 canonical ensemble \cite{YORK}.
\par
The ultimate form of the Hamiltonian operator, ${\hat H}$, will depend on
how one  chooses to define and then foliate  
the accessible spacetime \cite{BPP}. 
For example, 
 previously  in a 1+1-dimensional context \cite{abelian1},  
Kunstatter and the current author
chose  boundary conditions that  generalized  the spherically symmetric
(four-dimensional) formalism of Louko and Whiting \cite{LW}.
Technically speaking, the spacetime was foliated, from the
black hole horizon to the surface of the box, into spacelike
hypersurfaces. Moreover, each of these spatial slices
was constrained to approach the bifurcation point (i.e., the
horizon point of vanishing Killing vector) along a static
slice. These boundary conditions are a natural choice 
from a   thermodynamic perspective, inasmuch as
the resulting spacetime can be analytically continued
to the  Euclidean (black hole) instanton \cite{HH}.
\par
With these specified  conditions, the
Lorentzian Hamiltonian operator, ${\hat H}_{ex}$,\footnote{Following
\cite{MR}, we label
this Hamiltonian with the subscript $ex$ for
reasons that will soon be made clear.} was found to take on the
following form \cite{abelian1}:
\be
{\hat H}_{ex}={\hat H}_{+}-{\hat H}_{-},
\label{302}
\ee
where:
\be
{\hat H}_{+}= {\sqrt{-g_{tt}^+ J(\phi_+)}\over lG}
\left[1-\sqrt{1-{2lG{\hat M}\over J(\phi_+)}}\right],
\label{303}
\ee
\be
{\hat H}_- = {N\over G}\phi_o({\hat M})= {N\over G} J^{-1}(2lG{\hat M}).
\label{304}
\ee
Here, ${\hat H}_{\pm}$ is the  contribution to
the Hamiltonian from the outer/inner
boundary (i.e., box surface/horizon),
$g_{tt}^+$ is the time-time component of the metric
at the outer boundary, ${\hat M}$  corresponds to the mass operator,
 and $N$ is defined according to:
\be
N=\left.{d\over dx}\left[\sqrt{{-g_{tt}\over g_{xx}}}\right]
\right|_{\phi=\phi_o}.
\label{305}
\ee
\par
Taking the limit of an infinitely sized box (i.e., $\phi_+\rightarrow
\infty$)\footnote{Note that $J(\phi)$ is expected to increase
monotonically with $\phi$   for virtually any model that
admits black hole solutions \cite{CADY1}; so that, typically,
$J(\phi_+\rightarrow\infty)\rightarrow\infty$.} 
and fixing the  time coordinate of a boundary observer
so that $|g^+_{tt}/J(\phi_+)|\rightarrow 1$ at infinity,
we  obtain the anticipated result:
\be
{\hat H}_+\rightarrow {\hat M}. 
\label{306}
\ee
Here, we have chosen  a gauge
that is quite natural, as it fixes the coordinate time
equal to the proper time for an asymptotic observer
whose ``physical'' metric is an asymptotically flat one.
Meanwhile, Euclidean thermodynamic
considerations can be used to fix the value of $N$ such that
$N=2\pi/\beta$ \cite{abelian1}. Therefore, 
$H_- =2\pi\phi_o/G \beta=S_o/\beta$ (cf. Eq.(\ref{12})),
and so:
\be
{\hat H}_{ex}\rightarrow {\hat M}- {S_o\over\beta},
\label{307}
\ee
which  is precisely the (Helmholtz) free energy
of the system.
\par
We have included the label $ex$ on the Hamiltonian to
express that only the {\it exterior} region  of the
black hole spacetime (i.e., the region outside
of the horizon) has been considered with the above choice
of boundary conditions.
Such a  restriction is known, at least for spherically
symmetric gravity, to induce a free-energy form
for the Hamiltonian \cite{BPP}, in agreement
with our above (and prior \cite{abelian1}) findings.
\par
 Alternatively, one could have considered
a foliation that covers the entire black hole spacetime,
from left-hand-side to right-hand-side asymptotic
infinity (including the interior regions),  by considering
the maximally extended ``Kruskal'' diagram.\footnote{In our case,
this implies a suitable generalization of Kruskal coordinates \cite{KRU} 
for 1+1-dimensional gravity. Such coordinates have, indeed, been
explicitly formalized in \cite{THESIS}. Also note that
the ``left-hand side'' of the spacetime corresponds to the
submanifold  $x\leq 0$ (in which case, $\phi=-x/l\geq 0$).}
It is also of interest to know what the form of the Hamiltonian
would be for this latter choice of boundary conditions.
This form can, in fact, be
 readily extrapolated from  analogous calculations in the context of
 spherically symmetric gravity \cite{KUCHAR,BPP,MR}.  
Fixing the time coordinate at the right boundary
as above and freezing time evolution
at the left boundary,\footnote{The physical justification being
that an observer  should only be capable of making observations
at one asymptotic boundary.} we have in the asymptotic limit:
\be
{\hat H}_{wh}\rightarrow{\hat M},
\label{308}
\ee
where the subscript  $wh$ indicates that the {\it whole} black hole
spacetime has now been foliated. 
Evidently, with this choice of boundary conditions,
the Hamiltonian corresponds to  the internal energy of the
system. 
\par
On an intuitive level, one can understand the apparent
discrepancy between  Eq.(\ref{307})
and Eq.(\ref{308}).  If the horizon is to be taken seriously,
as a causal barrier for a fiducial observer,\footnote{That is,  a
static observer that remains outside of the horizon.} then such
an observer must necessarily be deprived of information
about the black hole interior. It is a general principle that any loss of 
information will directly translate
into a gain in entropy. (In the current discussion, this would be
an entanglement entropy from a quantum mechanical 
viewpoint.) This effect  is naturally manifested in our formalism
via the entropic contribution, ${\hat H}_-$, to the
exterior Hamiltonian. Conversely, a hypothetical observer
with access to the entire spacetime  will assign no special
meaning  to the horizon and  will, therefore, be impervious
to its information-negating effects.
That is to say, for this privileged observer,  
${\hat H}_-$  must  effectively vanish. 
\par
Recalling that our  current objective is to calculate
the partition function (\ref{301}), we must somehow deal with
the bothersome issue of having (at least) one Hamiltonian 
too many. On this point, we will argue
that, with proper handling, the partition function
describes the same thermodynamics, regardless
of whether one chooses to work with ${\hat H}_{ex}$ or 
${\hat H}_{wh}$. (As would naturally be expected from
any proponent of  black hole complementarity \cite{COMP}.)
 To support this claim, we will follow
the logistic framework of Makela and Repo \cite{MR}; which is based,
in large part, on  Bekenstein's  notion of black hole
spectroscopy \cite{BEKX}.
\par
We begin here by considering, in view of the arguments of 
Bekenstein and others \cite{BEKX,DASGAB},  a discrete spectrum for the mass
operator of the black hole.  Quantitatively speaking, one might
expect the following eigenvalue equation:
\be
{\hat M}|M_n> = M_n|M_n>, \quad\quad n=0,1,2,...,
\label{309}
\ee
where the mass eigenvalues, $M_n$, increase monotonically 
with  the quantum number $n$.  We will also
assume a level-dependent degeneracy,
which will be denoted by
$g_{ex}(M_n)$ or $g_{wh}(M_n)$. 
\par
Let us concentrate, for the moment, on ${\hat H}_{wh}={\hat M}$.
Directly incorporating this Hamiltonian and
the above formalism, we can  re-express
the partition function (\ref{301}) as follows:
\bea
{\cal Z}_{wh}[\beta]&=&\sum_{n=0}^{\infty}
<M_n|g_{wh}({\hat M})e^{-\beta {\hat M}}|M_n>
\nonumber\\
&=&\sum_{n=0}^{\infty}g_{wh}(M_n)e^{-\beta M_n}. 
\label{310}
\eea
We require, at this point,   some explicit form of
the degeneracy function.  
From the perspective of statistical mechanics,
 it is most natural to assume that
$\ln g_{wh}(M_n)=S_o(M_n)$, where $S_o$ is the classical black hole
entropy (\ref{12}). Also employing Eq.(\ref{100}),
we thus have:
\be
{\cal Z}_{wh}[\beta]=\sum_{n=0}^{\infty}\exp\left(-\beta M_n + 
{2\pi\over G}J^{-1}
(2lGM_n)\right). 
\label{311}
\ee
\par
Let us next consider ${\hat H}_{ex}$ as given by 
Eqs.(\ref{302},\ref{304},\ref{306}). 
With this choice of
Hamiltonian, the partition function (\ref{301}) now translates
into:
\be
{\cal Z}_{ex}[\beta]=\sum_{n=0}^{\infty}g_{ex}(M_n)\exp\left[-\beta \left(M_n-
{N\over G} J^{-1}(2lGM_n)\right)\right].
\label{312}
\ee
What about the degeneracy function in this case?  Here,
 we will argue  that $g_{ex}(M_n)$ should be  unity (for all $n$)
 on the grounds of the black hole no-hair theorem \cite{BALD}. 
That is to say, an asymptotic observer who is restricted to
the exterior  should  know nothing about the internal
degrees  of freedom of the black hole (which are presumably
encoded within the degeneracy). Rather,  such an observer should
only  be able to account for the macroscopic parameters;
which, in this toy model, are exclusively restricted to the mass.
\par
Setting $g_{ex}(M_n)=1$ and also fixing $N=2\pi/\beta$
(as follows from
Euclidean thermodynamic considerations \cite{abelian1}), we find
that ${\cal Z}_{wh}$ and ${\cal Z}_{ex}$ are indeed  indistinguishable.
Thus, we are lead to conjecture  a 
``universal''  partition function of the following form:
\be
{\cal Z}[\beta]=\sum_{n=0}^{\infty}\exp\left(-\beta M_n + 
{2\pi\over G}J^{-1}
(2lGM_n)\right). 
\label{uni}
\ee 
\par
A  point of interest: if one would rather avoid
the use of Euclidean thermodynamics to fix $N$
(which may seem unsavory in view of our otherwise
Lorentzian framework), then
 black hole complementarity \cite{COMP} might have alternatively
  been employed for this very  purpose.
To put it another way, if one subscribes {\it a priori}  to 
observer-independent physics,  then 
$N$ can be constrained (by identifying the partition functions)
 independently of any other considerations.

\section{Quantum-Corrected Entropy}

With a single, well-defined expression for the  partition function
(\ref{uni}), we  are now  in a position  for a quantum calculation of
the black hole entropy. 
If we are going to proceed, however,  an 
explicit form for the mass eigenvalues, $M_n$, will first be required.
To resolve this matter,  let us first take note of 
Bekenstein's proposal \cite{BEKX} for the eigenvalues of the horizon 
area:\footnote{In later adaptations  (for instance,  \cite{DASGAB}),
there is often a zero-point term included as well. However,
for a sufficiently massive black hole, such a term 
is essentially irrelevant and can be safely neglected.}
\be
A_n=\epsilon n l_p^2, \quad\quad n=0,1,2,...,
\label{401}
\ee
where $\epsilon$ is a numerical constant of order unity and
$l_p$ is the Planck length. This expression, of course,
 strictly applies 
to a four-dimensional theory.  Nonetheless, in view of 
Eqs.(\ref{area},\ref{100}), 
the most obvious 
two-dimensional
analogue can be written as: 
\be 
J^{-1}(2lGM) =\epsilon n, \quad\quad n=0,1,2,...,
\label{402}
\ee
or perhaps more usefully as:
\be
M_n={1\over 2lG} J(\epsilon n), \quad\quad n=0,1,2,...,
\label{403}
\ee
where $\epsilon$ is a constant  of order $G$.
\par
For definiteness, let us now call upon the power-law  model
as stipulated by Eqs.(\ref{15},\ref{150}).
In this case:
\be
M_n= {\gamma\over a+1} {(\epsilon  n)^{a+1}\over 2lG}.
\label{404}
\ee
\par
Substituting the above result into Eq.(\ref{uni}), we
have for the (universal) partition function:
\be
{\cal Z}[\beta]=\sum_{n=0}^{\infty}\exp\left( 
-\beta {\gamma\over a+1} {(\epsilon  n)^{a+1}\over 2lG}
+{2\pi\over G}\epsilon n
\right).
\label{405}
\ee
It is quite evident that this summation converges,
provided that $a\geq 0$, as has been priorly assumed.
However, the case of $a<0$ could still be readily handled
via  Makela and Repo's approach \cite{MR}; that 
is,  calculating the partition function for
the black hole radiation rather than the black
hole {\it per se}. (This maneuver effectively
reverses the sign of the exponent.) In this way,
one can verify that our final outcome (for the entropy)
can be extended to any choice of $a$.
\par
Let us now assume that the black hole is 
  sufficiently  massive  so that the  summation can be
replaced by an integral:
\be
{\cal Z}[\beta]=\int_{0}^{\infty}dy\exp\left( 
-\beta {\gamma\over a+1} {y^{a+1}\over 2lG}
+{2\pi\over G}y
\right),
\label{406}
\ee
where an obvious change has been made in the integration variable.
\par
Without resorting to numerics, one can still 
proceed by employing  a saddle-point
type of approximation.
First of all, a straightforward calculation reveals an extremum in the
exponent  at $y=y_o$, where $y_o^a=4\pi l/\beta\gamma$.
Next, we can expand  about this extremal point  to obtain:
\be
{\cal Z}[\beta]\approx 
e^{{2\pi a\over G(a+1)}\left[{4\pi l\over\beta\gamma}\right]^{1\over a}}
\int_{0}^{\infty}dy\exp\left(-{a\pi\over \gamma}
\left[{\beta\gamma\over 4\pi l}\right]^{1\over a}(y-y_o)^2 
\right).
\label{407}
\ee
This expression  is readily integrated to yield:
\be
{\cal Z}[\beta]\approx 
e^{{2\pi a\over G(a+1)}\left[{4\pi l\over\beta\gamma}\right]^{1\over a}}
\sqrt{{\gamma\over 4a}}\left(4\pi l\over \beta\right)^{1\over
2a}\left[1+erf(\sqrt{\omega})\right],
\label{408}
\ee
where:
\be
\omega = {a\pi\over \gamma}
\left({4\pi l\over \beta \gamma}\right)^{1\over a}
\label{409}
\ee
and  $erf$ denotes the  error function.\footnote{To be precise:
$erf(\alpha)=2\int^{\alpha}_{0}d\xi\exp(-\xi^2)/\sqrt{\pi}$.}
\par
As implied above,
we are  specifically  focusing on a semi-classical regime; that is,
the black hole mass should be regarded as large 
relative to any fundamental scale
 (which 
is a typical convention  in studies of this nature).
  It is
clear from  Eqs.(\ref{16},\ref{17}) that, for
the model of interest, the temperature increases
monotonically with mass. Hence, this semi-classical regime 
can  be equally well portrayed as one of high temperature.
This means that, under current considerations,
 we are justified in  regarding $\beta$ as small
in units of   the fundamental length scale, $l$. 
For this reason, $\omega$ can be viewed as a very large
(dimensionless) quantity; thus implying  that
$erf(\sqrt{\omega})\approx 1$ 
(up to corrections that vanish in the limit of infinite temperature).
\par
Let us
consider
the standard formula for extracting entropy from a thermodynamic
partition function: 
\be
S= \left(1-\beta{\partial\over\partial\beta}\right)\ln {\cal Z}.
\label{410}
\ee
Substituting for the partition function (\ref{408})  and
keeping the semi-classical approximation in mind,
we eventually have:
\be
S = {2\pi\over G}\left( {4\pi l\over \beta \gamma} \right)^{1\over a}
+{1\over 2a}\ln\left({l\over \beta}\right) 
+{\cal O}\left[\left({\beta\over l}\right)^{1\over a}\right]
+constant.
\label{411}
\ee
\par
Let us remind ourselves that
 $\beta^{-1}$ is defined as the equilibrium temperature measured
by an asymptotic observer. This is simply the definition of
the Hawking temperature,\footnote{Recall from Section 3
that the time coordinate of the  boundary observer has been
fixed so that the ``physical'' metric is asymptotically flat.
Therefore, as far as an asymptotic observer is concerned,
 the Tolman red-shift factor is equal to  unity.  Hence,
the asymptotic boundary temperature  does indeed coincide with the
Hawking value.}
 and so  it follows that $\beta^{-1}=T_o$.  
With this identification, 
we can apply Eq.(\ref{17}) for $T_o$, along with
 Eq.(\ref{12}) for the classical entropy ($S_o$), and
re-express Eq.(\ref{411}) as follows: 
\be
S =  S_o +{1\over 2}\ln(S_o)
+{\cal O}\left[ S_o^{-1}\right]
+constant.
\label{412}
\ee
\par
It is a nice consistency check that we have reproduced
the classical thermodynamic value with the lowest-order term. (Even
if this  was, more or less,  pre-ordained by  some
assumptions made in Section 3.)  A substantially less trivial
 outcome is, however,  the first-order quantum correction.
Here, we have confirmed the expected logarithmic dependence
and, moreover, found a prefactor of +1/2. It is interesting
and perhaps disturbing that this prefactor disagrees,
in both sign and magnitude, with the values obtained
(for the very same model)
 in prior calculations
\cite{MEME}. We will have more to say on this discrepancy
in the concluding section.

\section{Conclusion}

In summary, we have  been investigating
 a  generic theory of 1+1-dimensional
dilaton-gravity; with special attention  on
models with a power-law form of potential. 
After some preliminaries, we employed a
Lorentzian Hamiltonian approach as a means for
studying  the thermodynamics of two-dimensional
black holes.
 Black hole spectroscopy
also played a substantial role in establishing a  calculable
expression for the quantum  partition function.  Ultimately, we
were able to derive  an expression for the entropy
as an expansion in diminishing powers of  Hawking temperature.
(This temperature can be regarded as  large  
in a  semi-classical
regime of massive black holes.)  Reassuringly, 
the tree-level term   was shown to be in precise
agreement with the classical thermodynamic entropy.
Moreover, we have substantiated that the leading-order
correction  is directly proportional to the logarithm of
the classical entropy. However, the prefactor for
this logarithmic correction, +1/2, is in conspicuous disagreement
with the values obtained  (for the exact same model)
by other methodologies \cite{MEME}.
\par
 An interesting feature of our formalism (specifically in Section
3) is that it provides a simple but powerful framework
for understanding  how black hole complementarity \cite{COMP}
can resolve the information loss paradox \cite{PRES}.
As far as a  fiducial observer is concerned, the black hole
horizon is  a physical, causal boundary  behind which information can
indeed be lost. This apparent loss of information 
manifests itself  as an entanglement entropy
(between the interior and exterior regions) and  is
directly evident as an entropic contribution to the free energy
(cf. Eq.(\ref{307})). Conversely, from a global perspective,
the horizon is merely an artifact of poorly chosen coordinates, and 
there can be no loss of information 
in  this frame of reference. Rather, the black hole entropy
arises from an inherent degeneracy in the discrete levels of
the mass or area  spectrum. (Meanwhile, the fiducial observer
has no knowledge of this degeneracy by virtue of
the no-hair theorem.) That these two  viewpoints have
turned out to be
thermodynamically equivalent can be viewed as a manifestation
of black hole complementarity. (For further discussion,
see \cite{MR}.) 
\par
Finally, let us more closely examine the apparent
discrepancy that has arisen at the logarithmic order
in the  entropy. To review \cite{MEME}, a thermodynamic calculation
(based on \cite{das})
yields  a  prefactor of $-{1\over 2}(2a+1)$, a statistical
approach  (based on \cite{carlip}-\cite{sol}) 
gives a value of $-3/2$,
while the current, Hamiltonian program generates a value of $+1/2$.
\par
Assuming the  credibility of all three methodologies, 
one is inclined to take an inventory of any possible limitations.
In this regard, it is perhaps  relevant that the statistical
approach is based on a near-horizon duality and 
only accounts for  degrees of freedom within the vicinity
of the horizon. However, there is strong support
for the notion that this is precisely where the degrees of freedom
 of a black hole spacetime should reside (for instance, \cite{BW}).
On the other hand, we should re-emphasize that {\it both} the thermodynamic
and  Hamiltonian  method invoke an assumption of
large temperature. For the two-dimensional model of interest,
this regime of high temperature can best be viewed as a
semi-classical approximation. Since we are considering a
{\it quantum} correction  to the entropy, it
is quite possible that this regime is too restrictive.
On this basis,  one might argue that, in a true quantum  analysis,
the thermodynamic and Hamiltonian prefactors will both
converge towards the statistical value of $-3/2$.
\par
 Further support
that the statistical value of the prefactor is, in some sense, the fundamental
one follows from calculations in a quantum-geometry context. 
That is to say, a viable  candidate for a quantum theory of  gravity
also yields a value of $-3/2$ \cite{kaul}.
It may also be pertinent that the statistical approach is
based directly  on the Cardy formula \cite{car}.  More
to the point, 
it has been argued
that, exclusively for a two-dimensional theory,  the Cardy formula 
counts the true  microstates of the  black hole
when it is used in this manner
 \cite{fur}.
If this is literally correct,
the statistical calculation  would indeed be entitled to
an elevated status.
\par
On the other hand, it could be argued that the  various methods
are measuring independent contributions to the
first-order entropy, which should, at the end of the day, be summed
up. For instance, it has  recently been suggested \cite{NEW4}
that a positive prefactor should appear in calculations
that directly measure  fluctuations
in the horizon area, whereas a negative prefactor
should appear in calculations that count the number of
microstates.  Our results for the  Hamiltonian and statistical
methods do indeed appear to comply with this notion.
It does not, however,   seem {\it a priori} clear as to how the thermodynamic
approach should be classified in this scheme; although
the prefactor is, in this case, typically negative.
 \par
It would be interesting if any of  the above  arguments could
be supported (or disputed) in a more rigorous way,
and we hope to address this matter in a future study.

\section{Acknowledgments}

The author would like to thank V.P. Frolov for
helpful conversations.

\end{document}